\documentclass[aps,10pt,pra,twocolumn]{revtex4-1}
\usepackage{amsmath}
\usepackage{amssymb}
\usepackage{amsfonts}
\usepackage{relsize}
\usepackage{esint}
\usepackage{bbold}
\usepackage{graphicx}
\usepackage{dcolumn}
\usepackage{bm}
\usepackage{epsfig}
\usepackage{float}
\usepackage[colorlinks=true, citecolor=red, urlcolor=blue, linkcolor=blue]{hyperref}

\usepackage{bookmark}

\usepackage{hyperref}
\usepackage{cleveref}

\renewcommand{\vec}[1]{\boldsymbol{#1}}

\begin{document}

\title{Universal and robust quantum coherent control based on a chirped-pulse driving protocol}

\author{Yue-Hao Yin}
\affiliation{Center of Theoretical Physics, College of Physics, Sichuan University, Chengdu 610065, China}

\author{Jin-Xin Yang}
\affiliation{Center of Theoretical Physics, College of Physics, Sichuan University, Chengdu 610065, China}

\author{Li-Xiang Cen}
\email{lixiangcen@scu.edu.cn}
\affiliation{Center of Theoretical Physics, College of Physics, Sichuan University, Chengdu 610065, China}

\begin{abstract}
We propose a chirped-pulse driving protocol and reveal its exceptional property for quantum coherent control.
The nonadiabatic passage generated by the driving protocol, which includes the population inversion and
the nonadiabaticity-induced transition as its ingredients, is shown to be robust against pulse truncation.
We further demonstrate that the protocol allows for universal manipulation on
the qubit system through designing pulse sequences with either properly adjusted sweeping frequency or
pulsing intensity.
\end{abstract}

\maketitle

\section{Introduction}
\label[section]{intro}

Exquisite control over the dynamics of quantum systems is highly sought after
in various fields of quantum physics and engineering, including atomic
interferometry \cite{weitz1994,butts2013,kumar2013},
quantum-limited metrology \cite{giov2006,boixo2007,yang2022}, information processing
for qubit systems \cite{nielsen}, and so on.
In particular, achieving quantum gate operations
with sufficient accuracy that surpass the error thresholds of quantum error correcting
codes \cite{knill2005,preskill2006} is a crucial aspect in realizing scalable quantum computation.
A variety of fault-tolerant techniques, e.g., geometric quantum manipulation
\cite{zanardi1999,jones2000,duan2001,toyoda2013,review2023},
dynamically corrected gates \cite{dcg1,dcg2,dcg3} as well as numerical optimization \cite{kha2005,song2022},
have been proposed to implement coherent manipulation
for quantum states and information processing, aiming to address imperfections in fabrication
or the decoherence induced by the environmental noise.

Utilization of chirped pulses in driven quantum systems is able to produce robust state
transfer, which has been incorporated into quantum control schemes such as rapid adiabatic
passage \cite{melinger1992,vitanov2001,netz2002} and the composite pulse sequences
\cite{brown2004,toro2011}. Comparing with the conventional resonant
$\pi$-pulse scheme \cite{pi2002}, the adiabatic passage of the chirped-pulse driving
offers the advantage of being insensitive to the pulse area. On the other hand, the occurrence of avoided level
crossings in such driven systems can exhibit diverse dynamical behavior related
to the nonadiabatic evolution.
For example, the
adiabatic population transfer would be damaged by the
nonadiabaticity-induced transition, e.g., in the well-known Landau-Zener model \cite{landau,zener},
whereas this state transferring will be retained under the nonadiabatic evolution in some of its variants
\cite{yang2018,li2018}.
Moreover, chirped pulses assume an ideal infinite field intensity, which often results
in the generation of a divergent dynamical phase. In realistic systems, field pulses
are inevitably truncated at the starting and ending points. While this truncation may not
significantly impact the fidelity of the wave function, it does present challenges in accurately
controlling the phase factor. This, in turn,
will affect the coherent dynamics during subsequent information processing.

In this paper, we propose a distinctive chirped-pulse driving protocol and demonstrate
that its coherent dynamics is immune to the aforementioned pulse truncation.
The key to this property lies in
the fact that the total phase integrated over the nonadiabatic evolution of the driven
quantum system converges to a finite value, despite the infinite chirped field. Consequently,
we are able to reveal that the nonadiabatic passage, including the population inversion
and the nonadiabatic transition generated by the dynamical evolution,
is insensitive to truncation of the chirped pulse. Furthermore, we illustrate that this driving
protocol enables universal manipulation on the qubit system by designing a pulse sequence
with appropriately tuned frequency or field strength of the chirped pulses.

The remaining sections of the paper are organized as follows. In Sec. \ref{Solution}
we propose a chirped-pulse driven model and present an analytic approach to
resolving exactly its nonadiabatic evolution governed by the time-dependent Schr\"{o}dinger
equation. The time evolution operator of the ideal driving protocol
is shown to incorporate two elements: the population inversion and the
nonadiabaticity-induced transition.
Its explicit form is then elucidated in relation to different settings of the field parameters.
In Sec. \ref{truncation}, we reveal the robustness of the resulting coherent operation
of the protocol in the presence of truncation of the chirped pulse. Moving forward to Sec. \ref{QM},
we demonstrate how universal qubit manipulation can be achieved through designing
a pulse sequence with adjustable field parameters.
Finally, Sec. \ref{conclusion} provides a summary of the paper.

\section{Driven quantum model of the protocol and its exact solution}
\label[section]{Solution}

\begin{figure}[t]
	\includegraphics[width=0.9\columnwidth]{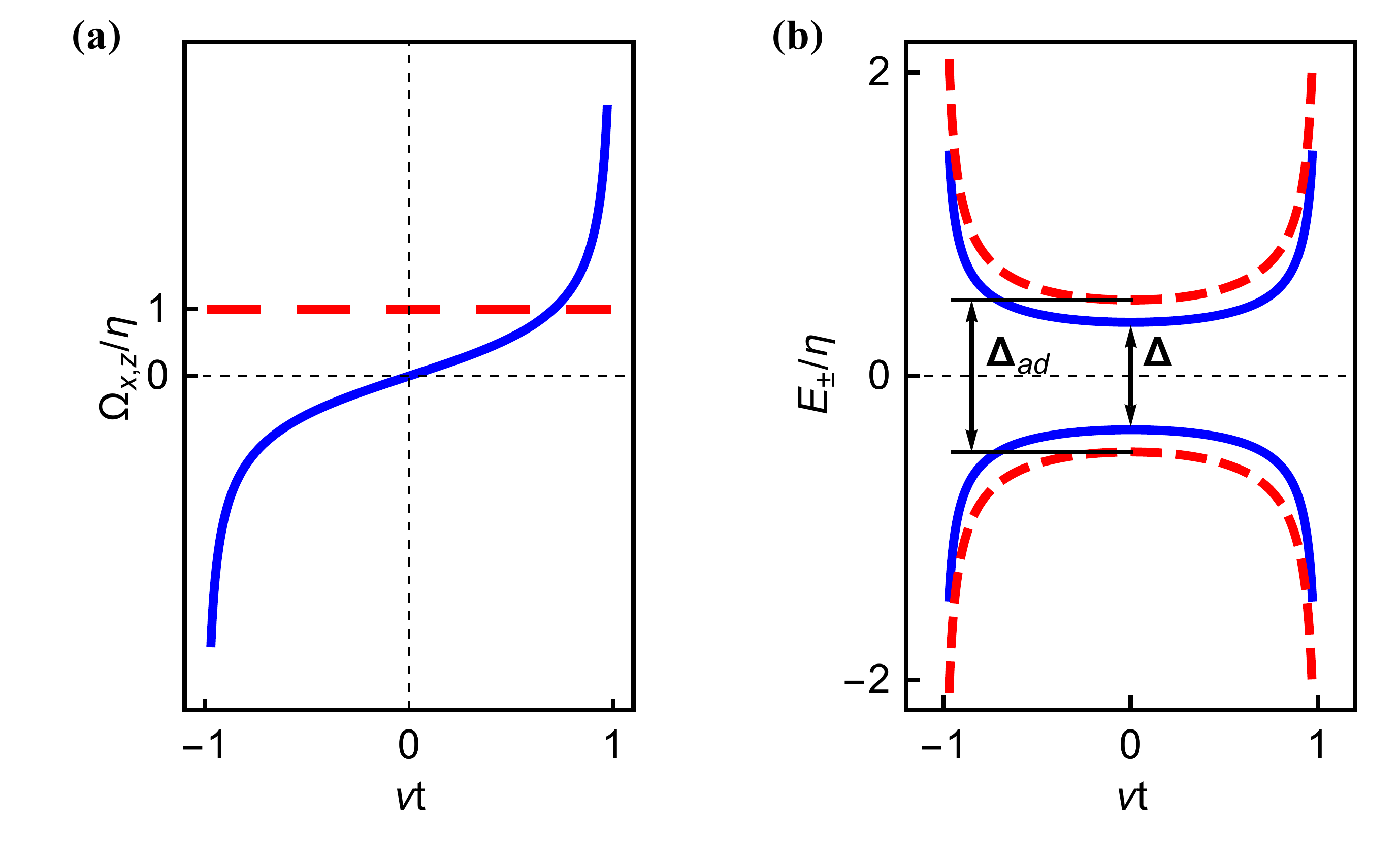}
	\caption{Chirped-pulse driving
protocol specified by the Hamiltonian (\ref{Hamil}).
(a) Field component $\Omega_z(t)$ (solid line) and $\Omega_x$ (dashed line)
of the driving protocol;
(b) The nonadiabatic energy described by $E_\pm(t)/\eta=\mp\frac 12\cos\varphi\csc\theta$ with $\nu/\eta = 1$
(solid line) and adiabatic energy levels in the limit $\nu/\eta\rightarrow 0$ (dashed line).
The corresponding gaps at the crossing point $t=0$
are shown to be $\Delta= 1/\sqrt{2}$ and $\Delta_{ad}=1$, respectively.
}
\label[figure]{f_Omega_E}
\end{figure}

The driven quantum model of the chirped-pulse protocol that we are going to consider is described
by the following time-dependent Hamiltonian
\begin{equation}
  H(t)=\vec{\Omega}(t)\cdot\vec{J}= \eta \left[J_{x} + \frac{\nu t}{\sqrt{1-(\nu t)^2}}J_z\right],
  \label{Hamil}
\end{equation}
where $J_i$ $(i=x, y, z)$ are angular momentum operators satisfying
$[J_i, J_j]=i\epsilon_{ijk}J_k$, and the amplitude $\eta$ and the frequency $\nu$ are
given constants. The $z$ component of the radical-form scanning field $\vec{\Omega}(t)$,
$\Omega_{z}(t)=\frac{\eta\nu t}{\sqrt{1-(\nu t)^2}}$, varies from $-\infty$ to
$+\infty$ during $t\in(-1/\nu,1/\nu)$ (assuming $\nu>0$ herein), while
its $x$ component remains constant over time: $\Omega_x=\eta$.
As the instantaneous energy levels undergo
an avoided crossing (see Fig. \ref{f_Omega_E}),
the system specifies a distinct example from those known paradigms such as
the Landau-Zener model \cite{landau,zener} and tangent-pulse
driven model \cite{yang2018}. Note that this radical-pulse driven model (\ref{Hamil})
as well as the latter two driving protocols have ever
been used in the shortcut-to-adiabatic method \cite{bason2012,malossi2013,stefanatos2019},
where their adiabatic evolution is employed as the target trajectories to
achieve the population transfer. Interestingly, we shall show that the
nonadiabatic evolution of this particular driven model can be rigorously resolved, which can be
utilized to achieve universal quantum coherent manipulation.

We now show that the wave function of the system governed by the
Schr\"{o}dinger equation (setting $\hbar = 1$)
\begin{equation}
i\frac{\partial}{\partial t}| \psi (t)\rangle = H(t) |\psi (t)\rangle
\label{schro}
\end{equation}
can be solved analytically.
To this end, we invoke a so-called gauge transformation \cite{wang1993,cen2003,ding2010}
$G(t)=e^{i\theta(t) J_y}e^{i\varphi J_x}$, in which the angle
$\varphi = \arccos\frac{\eta}{\sqrt{\eta^2+\nu^2}}$ and
\begin{equation}
\theta (t)=-\arccos\frac {\Omega_z(t)}{\Omega (t)}= -\arccos(\nu t)
\label{angle}
\end{equation}
with $\Omega(t)\equiv|\vec{\Omega}(t)|$. The transformed
state $|\psi^g(t)\rangle =G^\dagger(t)|\psi(t)\rangle$ is verified to
satisfy a covariant Schr\"{o}dinger equation
$i\frac{\partial}{\partial t}|\psi^g (t)\rangle = H^g(t)|\psi^g (t)\rangle$
where the effective Hamiltonian reads
\begin{equation}
  H^g (t) = G^\dagger H(t) G(t)-iG^\dagger\partial_tG(t)=\sqrt{\frac{\eta^2+\nu^2}{1-(\nu t)^2}}J_z.
\end{equation}
That is to say, in the new representation with respect to the transformation $G(t)$,
the system possesses a ``stationary" solution
$|\psi_{\pm}^g(t)\rangle=e^{\mp i\Theta (t,t_0)}|\pm\rangle$, in which
$|\pm\rangle$ denotes the eigenstates of $J_z$
with the magnetic quantum number $m=\pm\frac 12$ and the total phase $\Theta(t,t_0)$ can be rigorously
calculated as \begin{equation}
  \begin{split}
    \Theta(t,t_0)&=\frac 12\int_{t_0}^t\sqrt{\frac{\eta^2+\nu^2}{1-(\nu t')^2}}dt'\\
    &=\frac12\sqrt{1+\frac{\eta^2}{\nu^2}}\bigg[\arcsin(\nu t)-\arcsin(\nu t_0)\bigg].
  \end{split}
  \label{tphase}
\end{equation}
Consequently, the basic solution to the original Schr\"{o}dinger equation
is obtained as
\begin{equation}
  | \psi_\pm (t)\rangle=G(t)| \psi^g_\pm (t)\rangle=e^{\mp i\Theta(t,t_0)}e^{i\theta (t)J_y}e^{i\varphi J_x}|\pm\rangle,
  \label{solu}
\end{equation}
by which the nonadiabatic energy levels of the system, defined by
$E_\pm(t)\equiv \left\langle \psi_\pm(t)| H(t)| \psi_\pm(t)\right\rangle$,
are worked out to be $E_\pm(t)=\mp\frac {\eta} {2}\cos\varphi\csc\theta$.
The avoided crossing phenomena of these nonadiabatic levels
as well as the eigenvalues of $H(t)$ are depicted in Fig. \ref{f_Omega_E}(b).

Following the above result, the evolution operator of the system
over any time interval $t\in(t_0,t_f)$ is obtained straightforwardly as
\begin{equation}
  U(t_f,t_0) = G(t_f)U^g(t_f,t_0)G^\dagger(t_0),
  \label{U_orirep}
\end{equation}
in which
$U^g(t_f,t_0)=\exp[-i\Theta(t_f,t_0) J_z]$ denotes the one generated by $H^g(t)$ in the
stationary representation. For the ideal overall evolution during
$t\in(-1/\nu,1/\nu)$,
one has $\theta(t_0)=-\pi$ and $\theta(t_f)=0$. Hence the evolution operator reads
\begin{eqnarray}
 U_0(\eta/\nu)&\equiv &U(1/\nu,-1/\nu)  \nonumber \\
 &=&e^{i\varphi J_x}e^{-i2\Theta_0(\eta/\nu)J_z}e^{-i\varphi J_x}\times e^{i\pi J_y}
  \label{Uoper}
\end{eqnarray}
with $\Theta_0(\eta/\nu)=\sqrt{1+\frac {\eta^2}{\nu^2}}\frac \pi 2$.
The last factor $e^{i\pi J_y}\equiv i\sigma_y$ in the above expression of $U_0(\eta/\nu)$
indicates the population inversion induced by the chirped pulse.
The product of the first three factors can be re-expressed as
\begin{equation}
\tilde{U}_0(\eta/\nu)=e^{-i2\Theta_0(\eta/\nu)J(\varphi)},J(\varphi) \equiv\sin\varphi J_y+\cos\varphi J_z.
\label{nonad}
\end{equation}
It accounts for the nonadiabaticity-induced transition
and is verified to recover a pure phase shift over the basis states $|\pm\rangle$ in the adiabatic
limit $\nu/\eta\rightarrow 0$ (i.e., $\varphi\rightarrow 0$). Owing to the parameter-dependency
of $\tilde{U}_0(\eta/\nu)$, late on we will demonstrate that
universal manipulation on the qubit system can be implemented by a sequence of
pulsed operations generated by the protocol with a tunable $\eta/\nu$.
The latter fact distinguishes the present driving protocol from the previous tangent-pulse
protocol \cite{yang2018} as well as those based on
the transitionless algorithm \cite{bason2012,malossi2013,stefanatos2019}.

The above obtained expression for the evolution operator is applicable to any real parameter
$\eta$, but it is only valid for $\nu>0$. For a Hamiltonian formulated by Eq. (\ref{Hamil}) but
with $\nu\rightarrow -\nu$, its relation to the original one can be described by a
$\sigma_x$ flip: $H(t)\rightarrow \sigma_xH(t)\sigma_x$, where $\sigma_i$ ($i=x,y,z$) denotes
Pauli matrices. So
its generated evolution can be obtained by applying the same flip operation on $U_0(\eta/\nu)$:
\begin{equation}
U_0(\eta/\nu)\rightarrow \sigma_xU_0(\eta/\nu)\sigma_x=e^{i2\Theta_0(\eta/\nu)J(\varphi)}\times e^{-i\pi J_y}.
\label{evolu1}
\end{equation}
Moreover, if we perform further the transformation $\eta\rightarrow -\eta$ on the Hamiltonian with $-\nu$,
the result of Eq. (\ref{evolu1}) is still valid just noticing
that $\varphi \rightarrow \pi-\varphi$. At this stage, it is recognized that
the resulting unitary evolution herein indicates an inverse operation of the original
$U_0(\eta/\nu)$:
\begin{eqnarray}
&&~~~e^{i2\Theta_0(\eta/\nu)J(\pi-\varphi)}\times e^{-i\pi J_y}  \nonumber \\
&&=e^{-i\pi J_y}\times e^{i2\Theta_0(\eta/\nu)J(\varphi)}\equiv U_0^\dagger (\eta/\nu),
\label{evolu2}
\end{eqnarray}
where we have used the fact $J(\pi-\varphi)=e^{-i\pi J_y}J(\varphi)e^{i\pi J_y}$.
This result can also be verified in view that the transformation of the parameters
$(\eta,\nu)\rightarrow (-\eta,-\nu)$
indicates that $H(t)\rightarrow \sigma_zH(t)\sigma_z$, hence the evolution operator
changes as
$U_0(\eta/\nu)\rightarrow \sigma_zU_0(\eta/\nu)\sigma_z=U_0^\dagger(\eta/\nu)$.

\section{Robustness of the coherent operation against imperfect pulses}
\label[section]{truncation}

One remarkable property of the above driven model, which distinguishes it from
the Landau-Zener driving and other known analogs, is that despite the infinite
intensity of the chirped pulse, it generates a finite total phase. In addition
to the known advantage that population inversion is insensitive to the truncation
at the ending points $t_{0,f}=\mp\tau$ as long as $\Omega_z(\tau)\gg\Omega_x$,
this property also suggests that truncating the chirped pulses in the present model
may have a lesser impact on the accumulated phase $\Theta_0(\eta/\nu)$ and,
consequently, on the nonadiabaticity-induced transition specified by
$\tilde{U}_0(\eta/\nu)$ in Eq. (\ref{nonad}). That is to say, the coherent dynamics
generated by this driving protocol would be robust against
arbitrary truncation provided that $\Omega_z(\tau)/\Omega_x\gg 1$.

\begin{figure}[t]
	\includegraphics[width=0.9\columnwidth]{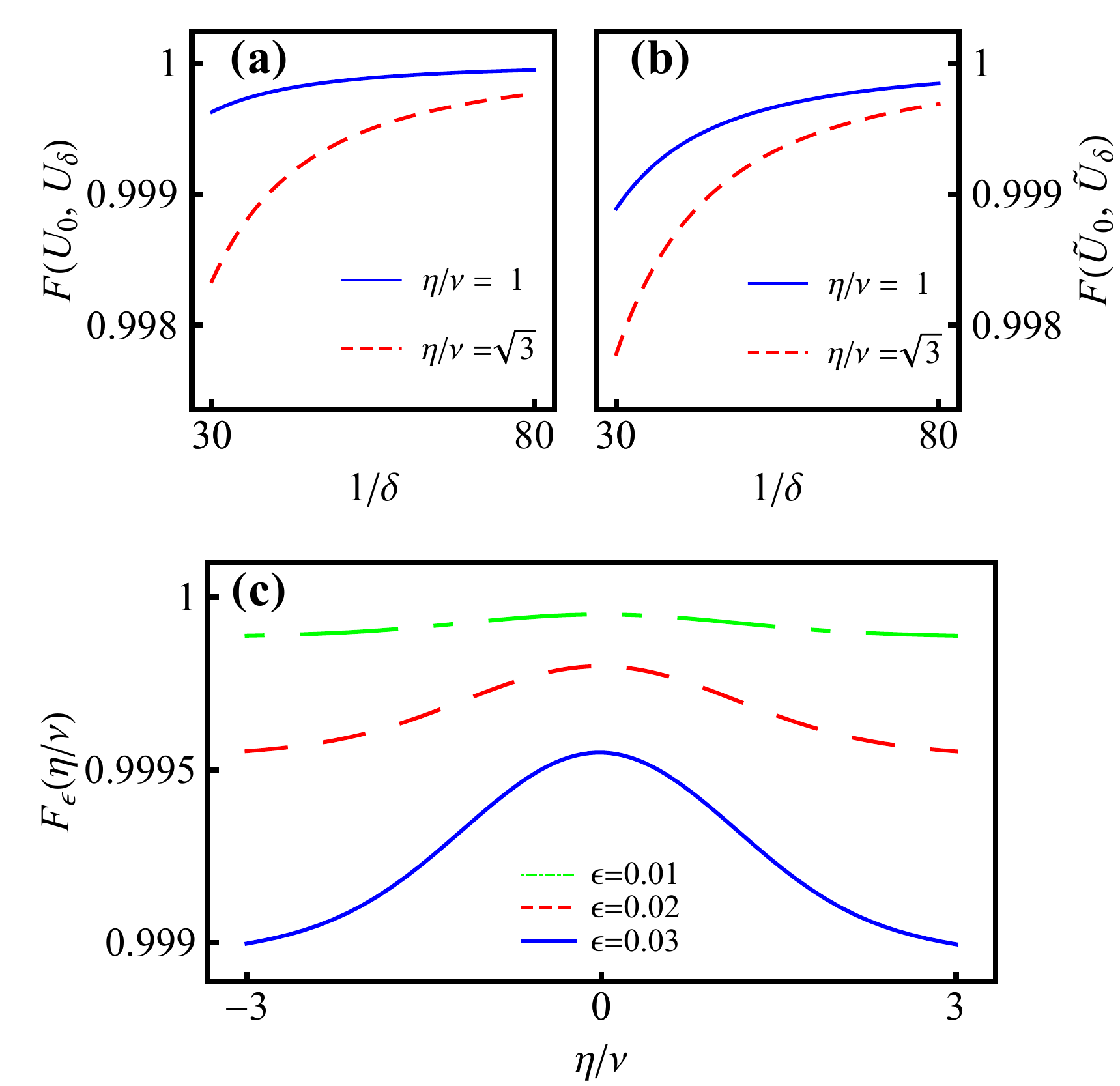}
	\caption{Fidelities of coherent operations yielded by the driving protocol
with pulse truncation. (a) $F(U_0,U_\delta)$ of the evolution operator
with $\eta/\nu = 1$ (blue solid line) and $\sqrt{3}$ (red dashed line);
(b) Fidelity between $\tilde{U}_0$ and $\tilde{U}_\delta$ accounting
for the nonadiabaticity-induced transition in the evolution operator;
(c) $F_\epsilon(\eta/\nu)$ between the desired evolution
$U_0(\eta/\nu)$ and the one $U_0(\tilde{\eta}/\tilde{\nu})$ with deviation of
the control parameter $\tilde{\eta}/\tilde{\nu}=\eta/\nu+\epsilon$.
}
\label[figure]{fidelity}
\end{figure}

To be more specific, let us denote by
$\delta \equiv\Omega_x/\Omega_z(\tau)$ the cutoff ratio of the field
components of the chirped pulse. Following Eq. (\ref{tphase}), the total phase
integrated over $t\in(-\tau,\tau)$ is given by
\begin{equation}
\Theta_\delta (\eta/\nu)=\sqrt{1+(\eta/\nu)^2}{\rm arccot}\delta,
\label{trphase}
\end{equation}
where we have used the relation $\nu\tau=(\delta^2+1)^{-1/2}$.
The resulting time evolution operator, according to Eq. (\ref{U_orirep}),
is specified by
\begin{equation}
 U_\delta(\eta/\nu)=e^{-i\arctan\delta J_y}\times\tilde{U}_\delta(\eta/\nu)\times e^{i(\pi-\arctan\delta)J_y},
  \label{evolucut}
\end{equation}
in which $\tilde{U}_\delta(\eta/\nu)=e^{-i2\Theta_\delta(\eta/\nu)J_\varphi}$
with $J_\varphi$ shown in Eq. (\ref{nonad}).
The influence of the truncation on the coherent dynamical evolution can be
conveniently characterized by the following fidelity:
\begin{equation}
F(U_0,U_\delta)=\frac 14 {\rm Tr}[U_0^\dagger U_\delta+U_\delta^\dagger U_0].
\end{equation}

Consider two particular settings of the dynamical parameter, say,
$\eta/\nu=\sqrt{3}$ ($\varphi=\pi/6$) and $\eta/\nu=1$ ($\varphi=\pi/4$).
For these two cases the total phases are worked out to be $\Theta_{0}(\eta/\nu)=\pi$ and $\pi/\sqrt{2}$,
and the ideal chirped-pulse driving gives rise to a spin flip $U_0(\sqrt{3})=e^{i\pi J_y}$ (along the $y$ axis)
and a composite flip operation $U_0(1)=e^{-i\pi(J_y+J_z)}e^{i\pi J_y}$, respectively.
For the practical driving process with pulse truncation, the corresponding evolution
operator $U_\delta (\eta/\nu)$
is given by Eq. (\ref{evolucut}) with
\begin{eqnarray}
\tilde{U}_\delta (\sqrt{3})&=&e^{-i2{\rm arccot}\delta(J_y+\sqrt{3}J_z)},  \nonumber \\
\tilde{U}_\delta (1)&=&e^{-i2{\rm arccot}\delta(J_y+J_z)}.
\end{eqnarray}
We plot in Fig. \ref{fidelity} the fidelities $F(U_0,U_\delta)$ of the above two coherent operations
with $\eta/\nu=\sqrt{3}$ and $1$ as well as $F(\tilde{U}_0,\tilde{U}_\delta)$ responsible
for those of the nonadiabaticity-induced transition. For all these quantities, the numerical results
display that the errors induced by the truncation could be reduced below the order of $10^{-3}$
as long as $\Omega_z(\tau)/\Omega_x \gtrsim 30$.

In addition to pulse truncation, practical implementations may also introduce
deviations in the control parameters. To account for the potential errors caused by these imperfections,
we evaluate the fidelity between the desired $U_0(\eta/\nu)$ and the one $U_0(\tilde{\eta}/\tilde{\nu})$
with deviation $\tilde{\eta}/\tilde{\nu}=\eta/\nu+\epsilon$, i.e.,
\begin{equation}
F_\epsilon(\eta/\nu)=
\frac 14 {\rm Tr}[U_0^\dagger(\eta/\nu) U_0(\tilde{\eta}/\tilde{\nu})+U_0^\dagger(\tilde{\eta}/\tilde{\nu})U_0(\eta/\nu)].
\end{equation}
The numerical result displays that the error is below $10^{-3}$ as long
as $\epsilon\lesssim 0.03$ [see Fig. \ref{fidelity} (c)].

\section{Universal qubit manipulation}
\label[section]{QM}

Generally, an arbitrary single-qubit operation can be implemented by
employing two non-commutative unitary transformations, e.g., the rotations
along the $y$ axis $e^{i\phi_y\sigma_y}$ and along the $z$
axis $e^{i\phi_z\sigma_z}$.
In the present protocol described by the Hamiltonian (\ref{Hamil}), while
non-commutative $U_0(\eta/\nu)$'s can be achieved by setting
different pulsing strength $\eta$ or sweeping frequency $\nu$,
they distinctly differ from those conventional rotations along fixed axes.
Therefore, it is interesting to inquire about the possibility and method of
implementing universal manipulation on the qubit system by the present driving
protocol. Below we provide an affirmative answer by outlining a pathway
to this goal.

Let us start by rewriting the time evolution operator shown in Eq. (\ref{Uoper}) as
\begin{equation}
U_0(\eta/\nu)=r_0I_2+i\vec{r}\cdot\vec{\sigma},
\label{bloch}
\end{equation}
where $r_0$ and $\vec{r}$ satisfy $r_0^2+\vec{r}^2=1$ and are specified by
\begin{eqnarray}
&&r_0(\eta/\nu)=\sin\Theta_0\sin\varphi, \nonumber \\
&&\vec{r}(\eta/\nu)=-\sin\Theta_0\cos\varphi \hat{e}_x+\cos\Theta_0\hat{e}_y.
\label{blovect}
\end{eqnarray}
We employ the evolution operators generated by a couple of consecutive driving pulses,
i.e., $U_0(\eta/\nu)$ and $U_0(\eta^\prime/\nu^\prime)=r_0^\prime I_2+i\vec{r}^\prime\cdot\vec{\sigma}$.
Combining these two operations gives rise to
\begin{equation}
\mathcal{R}(\eta^\prime/\nu^\prime,\eta/\nu)\equiv U_0(\eta^\prime/\nu^\prime)U_0(\eta/\nu)=R_0I_2+i\vec{R}\cdot\vec{\sigma},
\end{equation}
in which
\begin{eqnarray}
&&R_0=r_0^\prime r_0-r_x^\prime r_x-r_y^\prime r_y,~
R_x=r_0r_x^\prime +r_0^\prime r_x,   \nonumber \\
&&R_y=r_0r_y^\prime +r_0^\prime r_y,~~~~~~~~~~
R_z=r_xr_y^\prime -r_x^\prime r_y.
\end{eqnarray}

\begin{figure}[t]
	\includegraphics[width=0.9\columnwidth]{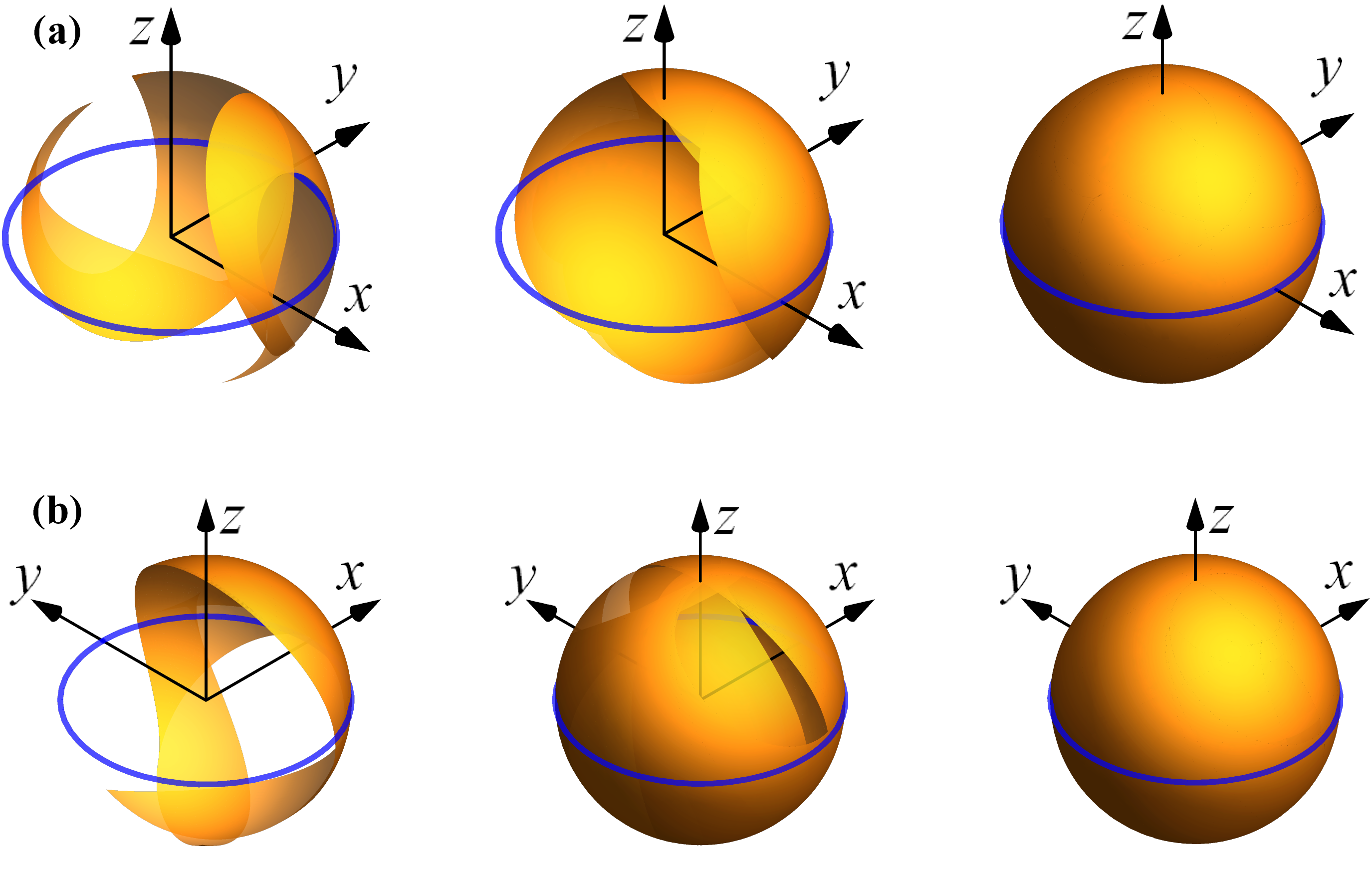}
	\caption{Schematic to illustrate the universality of the rotation $\mathcal{R}(\eta^\prime/\nu^\prime,\eta/\nu)$
acting on $\sigma_y$ and $\sigma_z$. (a) Parameter surfaces of $\vec{S}(\eta^\prime/\nu^\prime,\eta/\nu)$
in which the domains of $\eta/\nu$ and $\eta^\prime/\nu^\prime$ responsible for the left, middle, and right panels are
$\eta/\nu\in(-1,1),(-2,2),(-3,3)$ and $\eta^\prime/\nu^\prime\in(-1,1),(-2,2),(-3,3)$, respectively.
(b) Parameter surfaces of $\vec{T}(\eta^\prime/\nu^\prime,\eta/\nu)$ with respect to
the same domains of $\eta/\nu$ and $\eta^\prime/\nu^\prime$ specified in (a)
for the left to right panels.
}
\label[figure]{rotation}
\end{figure}

To proceed, we show the capability of the operation
$\mathcal{R}(\eta^\prime/\nu^\prime,\eta/\nu)$ to implement universal rotation
on a spin initially along either the $y$ or the $z$ axis.
Straightforwardly, we characterize the corresponding outcomes as
$\mathcal{R}\sigma_y \mathcal{R}^\dagger =\vec{S}\cdot\vec{\sigma}$ and
$\mathcal{R}\sigma_z \mathcal{R}^\dagger =\vec{T}\cdot\vec{\sigma}$,
in which the two Bloch vectors $\vec{S}(\eta^\prime/\nu^\prime,\eta/\nu)$
and $\vec{T}(\eta^\prime/\nu^\prime,\eta/\nu)$ are specified by
\begin{eqnarray}
&&S_x(\eta^\prime/\nu^\prime,\eta/\nu)=2R_0R_z+2R_xR_y, \nonumber \\
&&S_y(\eta^\prime/\nu^\prime,\eta/\nu)=R_0^2-R_x^2+R_y^2-R_z^2,   \nonumber \\
&&S_z(\eta^\prime/\nu^\prime,\eta/\nu)=-2R_0R_x+2R_yR_z,
\end{eqnarray}
and
\begin{eqnarray}
&&T_x(\eta^\prime/\nu^\prime,\eta/\nu)=-2R_0R_y+2R_xR_z, \nonumber \\
&&T_y(\eta^\prime/\nu^\prime,\eta/\nu)=2R_0R_x+2R_yR_z,   \nonumber \\
&&T_z(\eta^\prime/\nu^\prime,\eta/\nu)=R_0^2-R_x^2-R_y^2+R_z^2,
\end{eqnarray}
respectively. The parametric surfaces of $\vec{S}(\eta^\prime/\nu^\prime,\eta/\nu)$
and $\vec{T}(\eta^\prime/\nu^\prime,\eta/\nu)$
are shown in Fig. \ref{rotation}. The universality of the rotation is justified
by the fact that the function domains cover over the entire surface of
the Bloch sphere through tuning appropriately the ranges of
$\eta/\nu$ and $\eta^\prime/\nu^\prime$.

With the above results, we are now able to show that the universal set of the
gate operations $e^{i\phi_y \sigma_y}$ and
$e^{i\phi_z \sigma_z}$ with $\phi_{y,z}\in[-\pi/2,\pi/2]$
can be achieved by applying the driving protocol. Specifically,
one can exploit the following sequence of pulsed operations:
\begin{equation}
\mathcal{R}^\dagger(\eta^\prime/\nu^\prime,\eta/\nu)
U_0(\bar{\eta}/\bar{\nu})\mathcal{R}(\eta^\prime/\nu^\prime,\eta/\nu)\Rightarrow e^{i\phi_{y,z} \sigma_{y,z}}.
\label{sequ}
\end{equation}
Here, $U_0(\bar{\eta}/\bar{\nu})=\bar{r}_0I_2+i\bar{\vec{r}}\cdot\vec{\sigma}$ [cf. Eq. (\ref{bloch})]
denotes a ``source" operation generated by the driving protocol with a given
value of $\bar{\eta}/\bar{\nu}$. The universality of the rotation $\mathcal{R}(\eta^\prime/\nu^\prime,\eta/\nu)$
shown previously warrants that the Bloch vector $\bar{\vec{r}}$ can be transformed
along the $\pm y$ or $\pm z$ axis by the corresponding inverse transformation
$\mathcal{R}^{\dagger}(\eta^\prime/\nu^\prime,\eta/\nu)$. Consequently, the two gate operations
$e^{i\phi_{y,z} \sigma_{y,z}}\equiv \cos\phi_{y,z}I_2+i\sin\phi_{y,z}\sigma_{y,z}$
are obtained and the corresponding angle $\phi_{y,z}$, according to Eq. (\ref{blovect}),
is specified by:
\begin{equation}
\cos\phi_{y,z} =\sin\Theta_0 (\bar{\eta}/\bar{\nu})\sin\bar{\varphi},
\label{phi}
\end{equation}
where $\bar{\varphi}=\arccos\frac{\bar{\eta}}{\sqrt{\bar{\eta}^2+\bar{\nu}^2}}$
and we have used the fact that the coefficient of $I_2$ in $U_0(\bar{\eta}/\bar{\nu})$
is invariant under the rotation of Eq. (\ref{sequ}).
A simple numerical analysis is able to reveal that the value of $|\phi_{y,z}|$
covers the domain $[0,\pi/2]$ by tuning the ratio $\bar{\eta}/\bar{\nu}$ within
$0\leq\bar{\eta}/\bar{\nu}\leq\sqrt{3}$. As the sign of $\phi_{y,z}$, i.e., the orientation
of $e^{i\phi_{y,z} \sigma_{y,z}}$ along $\pm y$ and $\pm z$, can be freely
adjusted by the rotation $\mathcal{R}(\eta^\prime/\nu^\prime,\eta/\nu)$,
one can conclude that the pulse sequence of Eq. (\ref{sequ}) is able
to perform the promising universal gate operation.

\section{Conclusion}
\label[section]{conclusion}

In practical implementation, the spectral broadening due to the large amplitude
of the chirped pulse may result in leakage out of the computational qubit states.
Especially, this error may become dramatic for the transmon qubits \cite{RMP2021},
since their computational states are isolated by the weak anharmonicity and enhancing
the latter
(i.e., the charging energy) will lead to large dephasing rate.
At this stage, it is of interest to explore further the possible
way to incorporate the dynamical decoupling strategy \cite{bylander2011} into
the present driving protocol in order to mitigate these noise effects.

In summary, we have proposed a robust design for quantum coherent control
based on a particular chirped-pulse driving protocol.
The nonadiabatic passage induced by the driven model,
including the population inversion and the nonadiabaticity-induced transition associated with
dynamical evolution, is shown to be insensitive to the truncation
of the chirped pulse. Moreover, we illustrate that this driving protocol enables
universal manipulation of single-qubit systems by designing pulse sequences with
appropriately tuned frequencies or field strengths. Note that this universality of
local qubit operations, together with an arbitrary two-qubit interaction, is sufficient
to perform universal quantum computation \cite{dodd2002}. The simple and unified form of the driving
protocol undoubtedly will mitigate the intricacies with respect to the quantum information
processing hardware design. We therefore expect that this protocol holds significant
potential for physical realization, encompassing not only quantum coherent control
but also scalable quantum computation.

\acknowledgments {This work was supported by the National Natural
Science Foundation of China (Grant No. 12147207).}


\thebibliography{99}


\bibitem{weitz1994} M. Weitz, B.C. Young, and S. Chu, Atomic interferometer based on adiabatic population transfer. Phys. Rev.
Lett. {\bf 73}, 2563 (1994).


\bibitem{butts2013}  D.L. Butts, K. Kotru, J.M. Kinast, A.M. Radojevic, B.P. Timmons, and R.E. Stoner,
J. Opt. Soc. Am. B. {\bf 30}, 922 (2013).

\bibitem{kumar2013} P. Kumar and A.K. Sarma, Phys. Rev. A. {\bf 87}, 025401 (2013).

\bibitem{giov2006} V. Giovannetti, S. Lloyd, and L. Maccone, Phys. Rev. Lett. {\bf 96}, 010401 (2006).

\bibitem{boixo2007} S. Boixo, S.T. Flammia, C.M. Caves, and JM Geremia, Phys. Rev. Lett. {\bf 98}, 090401 (2007).

\bibitem{yang2022} J. Yang, S. Pang, Z. Chen, A.N. Jordan, and A. delCampo, Phys. Rev. Lett. {\bf 128} 160505 (2022).

\bibitem{nielsen} M.A. Nielsen and I.L. Chuang, Quantum Computation and Quantum Information (Cambridge University Press,
Cambridge, UK, 2000).

\bibitem{knill2005} E. Knill, Nature (London) {\bf 434}, 39 (2005).

\bibitem{preskill2006} P. Aliferis, D. Gottesman, and J. Preskill, Quantum Inf. Comput. {\bf 6}, 97 (2006).

\bibitem{zanardi1999} P. Zanardi and M. Rasetti, Phys. Lett. A {\bf 264}, 94 (1999).

\bibitem{jones2000} J.A. Jones, V. Vedral, A. Ekert, and G. Castagnoli, Nature
(London) {\bf 403}, 869 (2000).

\bibitem{duan2001} L.M. Duan, J.I. Cirac, and P. Zoller, Science {\bf 292}, 1695
(2001).

\bibitem{toyoda2013} K. Toyoda, K. Uchida, A. Noguchi, S. Haze, and S. Urabe,
Phys. Rev. A {\bf 87}, 052307 (2013).


\bibitem{review2023} J. Zhang, T.H. Kyaw, S. Filipp, L.C. Kwek, E. Sj\"{o}qvist,
D.M. Tong, Phys. Rep. {\bf 1027}, 1 (2023).

\bibitem{dcg1} K. Khodjasteh and L. Viola, Phys. Rev. Lett. {\bf 102}, 080501 (2009).

\bibitem{dcg2} K. Khodjasteh and L. Viola, Phys. Rev. A {\bf 80}, 032314 (2009).

\bibitem{dcg3} J. Zeng, C.H. Yang, A.S. Dzurak, and E. Barnes, Phys. Rev. A {\bf 99}, 052321 (2019).


\bibitem{kha2005} N. Khaneja, T. Reiss, C. Kehlet, T. Schulte-Herbr\"{u}ggen,
and S. J. Glaser, J. Magn. Reson, {\bf 172}, 296 (2005).

\bibitem{song2022} Y. Song, J. Li, Y.-J. Hai, Q. Guo, and X.-H. Deng, Phys.
Rev. A {\bf 105}, 012616 (2022).


\bibitem{melinger1992} J.S. Melinger, S.R. Gandhi, A. Hariharan, J.X. Tull, and W.S. Warren, Phys. Rev. Lett. {\bf 68}, 2000 (1992).


\bibitem{vitanov2001} N.V. Vitanov, T. Halfmann, B.W. Shore, and K. Bergmann, Annu. Rev. Phys. Chem. {\bf 52}, 763 (2001).


\bibitem{netz2002} R. Netz, T. Feurer, G. Roberts, and R. Sauerbrey, Phys. Rev. A {\bf 65}, 043406 (2002).



\bibitem{brown2004} K.R. Brown, A.W. Harrow, and I.L. Chuang, Phys.
Rev. A. {\bf 70}, 052318 (2004).

\bibitem{toro2011} B.T. Torosov, S. Gu\'{e}rin, and N.V. Vitanov, Phys. Rev.
Lett. {\bf 106}, 233001 (2011).


\bibitem{pi2002} U. Boscain, G. Charlot, J.-P. Gauthier, S. Gu\'{e}rin, and H.-R. Jauslin,
J. Math. Phys. {\bf 43} 2107 (2002).


\bibitem{landau} L.D. Landau, Phys. Z. Sowjetunion {\bf 2}, 46 (1932).

\bibitem{zener} C. Zener, Proc. R. Soc. A {\bf 137}, 696 (1932).


\bibitem{yang2018} G. Yang, W. Li, and L.-X. Cen, arXiv: 1608.00735; Chin. Phys. Lett. {\bf 35}, 013201
(2018).

\bibitem{li2018} W. Li and L.-X. Cen, Ann. Phys. (NY) {\bf 389}, 1 (2018).


\bibitem{bason2012} M.G. Bason, M. Viteau, N. Malossi, P. Huillery, E. Arimondo, D. Ciampini,
R. Fazio, V. Giovannetti, R. Mannella, and O. Morsch, Nat. Phys. {\bf 8}, 147 (2012).

\bibitem{malossi2013} N. Malossi, M.G. Bason, M. Viteau, E. Arimondo, R. Mannella, O. Morsch,
and D. Ciampini, Phys. Rev. A {\bf 87}, 012116 (2013).


\bibitem{stefanatos2019} D. Stefanatos and E. Paspalakis, Phys. Rev. A {\bf 100}, 012111 (2019).


\bibitem{wang1993} S.J. Wang, F.L. Li, and A. Weiguny, Phys. Lett. A {\bf 180}, 189
(1993).

\bibitem{cen2003} L.-X. Cen, X.Q. Li, Y.J. Yan, H.Z. Zheng, and S.J. Wang,
Phys. Rev. Lett. {\bf 90}, 147902 (2003).


\bibitem{ding2010} Z.-G. Ding, L.-X. Cen, and S.J. Wang, Phys. Rev. A {\bf 81},
032337 (2010).



\bibitem{RMP2021} A. Blais, A.L. Grimsmo, S.M. Girvin, and A. Wallraff,
Rev. Mod. Phys. {\bf 93}, 025005 (2021).

\bibitem{bylander2011} J. Bylander, S. Gustavsson, F. Yan, F. Yoshihara,
K. Harrabi, G. Fitch, D.G.Cory, Y. Nakamura, J.-S. Tsai, and W.D. Oliver, Nat. Phys. {\bf 7}, 566 (2011).



\bibitem{dodd2002} J.L. Dodd, M.A. Nielsen, M.J. Bremner, and R.T. Thew, Phys. Rev. A {\bf 65}, 040301(R) (2002).
\end{document}